\documentclass[%
 reprint,
 amsmath,amssymb,
 aps,
]{revtex4-2}

\usepackage{graphicx}
\usepackage{dcolumn}
\usepackage{bm}

\begin{document}

\preprint{APS/123-QED}

\title{First principles study of the T-center in Silicon}
\thanks{To whom correspondence should be addressed geoffroy.hautier@dartmouth.edu}%

\author{Diana Dhaliah}
\affiliation{%
 Institute of Condensed Matter and Nanosciences (IMCN), Universit\'{e} Catholique de Louvain, Chemin des \'{E}toiles 8, B-1348 Louvain-la-Neuve, Belgium
}%

\author{Yihuang Xiong}%

\affiliation{%
 Thayer School of Engineering, Dartmouth College, 14 Engineering Dr, Hanover, NH 03755, USA 
}%

\author{Alp Sipahigil}

\affiliation{
 Department of Electrical Engineering and Computer Sciences, University of California,  Berkeley, Berkeley, California 94720, USA
}%
\affiliation{
 Department of Physics, University of California, Berkeley, Berkeley, California 94720, USA
}%
\affiliation{
 Materials Sciences Division, Lawrence Berkeley National Laboratory, Berkeley, California 94720, USA
}%

\author{Sin\'{e}ad M. Griffin}
\affiliation{%
 Materials Sciences Division, Lawrence Berkeley National Laboratory, Berkeley, California 94720, USA
}%
\affiliation{%
 Molecular Foundry, Lawrence Berkeley National Laboratory, Berkeley, California 94720, USA
}%

\author{Geoffroy Hautier}

\affiliation{%
 Thayer School of Engineering, Dartmouth College, 14 Engineering Dr, Hanover, NH 03755, USA
}%
\affiliation{%
 Institute of Condensed Matter and Nanosciences (IMCN), Universit\'{e} Catholique de Louvain, Chemin des \'{E}toiles 8, B-1348 Louvain-la-Neuve, Belgium
}%

\date{\today}

\begin{abstract}
 The T-center in silicon is a well-known carbon-based color center that has been recently considered for quantum technology applications. Using first principles computations, we show that the excited state is formed by a defect-bound exciton made of a localized defect state occupied by an electron to which a hole is bound. The localized state is of strong carbon \textit{p} character and reminiscent of the localization of the unpaired electron in the ethyl radical molecule. The radiative lifetime for the defect-bound exciton is calculated to be on the order of $\mu$s, much longer than other quantum defects such as the NV center in diamond and in agreement with experiments. The longer lifetime is associated with the small transition dipole moment as a result of the very different nature of the localized and delocalized states forming the defect-bound exciton. Finally, we use first principles calculations to assess the stability of the T-center. We find the T-center to be stable against decomposition into simpler defects when keeping the stoichiometry fixed. However, we identify that the T-center is easily prone to (de)hydrogenation and so requires very precise annealing conditions (temperature and atmosphere) to be efficiently formed.
\end{abstract}

\maketitle


\section{Introduction}

There is a strong interest in controlling optically active point defects in semiconductors for applications in quantum communication~\cite{Pompili2021}, sensing~\cite{Hsieh2019}, and computing~\cite{Bradley2019,Yan2021}. These ``quantum defects" act as artificial atoms that can couple long-lived spin states and optical photons for optical spin readout and entanglement distribution over long distances~\cite{Atatre2018, Wolfowicz2021}. Well-studied quantum defects include the NV center in diamond, and the silicon divacancy in SiC, which have already demonstrated important initial steps for the development of quantum computers and sensors~\cite{Pompili2021}. However, silicon has significant potential benefits over diamond as a quantum defect host including its ease of scalable integration into photonic and electronic circuits~\cite{Durand2021,Yan2021,Saeedi2013}. In fact, color centers in silicon have a long history with numerous studies identifying and characterizing these defects prior to interest in quantum applications~\cite{Davies1989}. Defects made by carbon implantation have received much attention with a series of defects identified by letters from the alphabet (G-center, I-center, etc.). Of these, the T-center (composed of two carbons and one hydrogen) is of particular interest for  applications as it shows emission in the technologically relevant O-band of the infra-red spectra (1260 to 1360 nm) that is split under a magnetic field. This last requirement is critical as defects with unpaired electrons (i.e., having a doublet or triplet ground state) are necessary to store quantum information through spin. Though the T-center  was first reported in 1981 by Mianev and Mudryi~\cite{Minaev1981}, more recent work proposed to use the T-center for quantum applications using isotopically pure silicon~\cite{Bergeron2020,MacQuarrie2021}. The T-center has become an exciting candidate for a high-performance quantum defect in silicon that could become an key building block for future quantum technologies. 

First-principles calculations have proved essential for the understanding of quantum defects, including calculations of their relevant structural, electronic, optical, and magnetic properties~\cite{Ivady2018,Lee2019,Dreyer2018}. Here, we report on a first-principles study of the T-center. Previous first-principles results on the T-center focused mainly on vibrational frequencies and were carried out over a decade ago~\cite{Leary1998,Safonov1996}. However, major developments in \textit{ab initio} methodology and compute power over the past decade motivate us to revisit first principles calculations of the T-center to study its electronic structure in detail and identify the bound exciton nature of its excited state. We confirm and rationalize the rather long radiative lifetime ($\mu$s) measured experimentally. Finally, we analyze the thermodynamic stability of the T-center and identify dehydrogenation as the primary stability concern for this defect.

\section{Methods}

Our first-principles calculations were performed using VASP~\cite{G.Kresse-PRB96,G.Kresse-CMS96} and the projector-augmented wave (PAW) framework~\cite{P.E.Blochl-PRB94}. We used spin polarized computations with supercells that are relaxed at fixed volume until the forces on the ions are smaller than 0.01~eV/\r{A}. We used a 512 atoms supercell and a $\Gamma$-only k-point sampling. All computations were performed using the Heyd-Scuseria-Ernzerhoff (HSE)~\cite{Heyd2003} functional with 25\% exact exchange except when specified otherwise. The HSE functional provides an accurate band gap for silicon, 1.114 eV, which compares well with the measured bandgap of 1.17 eV at 0 K. More specifically, the decomposition and hydrogenation computations are performed using the generalized gradient approximation (GGA) with the Perdew-Burke-Ernzerhof (PBE) functional~\cite{J.Perdew-PRL96}. For all defect calculations, the input generation and output analysis were performed using PyCDT~\cite{Broberg2018}.
 The formation energy of each charged-defect state is calculated as a function of the Fermi level $E_f$ as~\cite{Zhang1991,Komsa2012}
\begin{equation}
    E_\mathrm{form}[X^q] = E_\mathrm{tot}[X^q] - E_\mathrm{tot}^\mathrm{bulk} - \sum n_i \mu_i + q E_f + E_\mathrm{corr}
\label{eq:defect}
\end{equation}
where $E_\mathrm{tot}[X^q]$ and $E_\mathrm{tot}^\mathrm{bulk}$ are the total energies of the defect-containing supercell (for a given defect $X$ in the charge state $q$) and the bulk, respectively. The third term represents the energy needed to exchange atoms with thermodynamic reservoirs where $n_i$ indicates the number of atoms of species $i$ removed or added to create the defect, and $\mu_i$ their corresponding chemical potential. The fourth term represents the energy to exchange electrons with the host material through the electronic chemical potential given by the Fermi level. Finally, the last term is a correction accounting for spurious image-charge Coulomb interactions due to finite supercell size, as well as potential-alignment corrections to restore the position of the bulk valence band maximum (VBM) in charged-defect calculations due to the presence of the compensating background charge density~\cite{Freysoldt2011,Kumagai2014}. The chemical potentials for silicon and carbon were set to the energies of their elemental solid.

We further computed the hyperfine coupling parameters using the relaxed 512 atoms supercell at HSE level with VASP. The hyperfine tensor $A_{ij}^I$ of nucleus $I$ comprises the isotropic Fermi-contact term and the anisotropic spin dipolar term, which are described by the two equations below, respectively:

\begin{equation}
{A_{\mathrm{iso}}^{I}}=\frac{2}{3} \frac{\mu_{0} \gamma_{e} \gamma_{I}}{S} \delta_{i j} \int \delta_{T}(\mathbf{r}) \rho_{s}\left(\mathbf{r}+\mathbf{R}_{I}\right) d \mathbf{r},
\end{equation}

\begin{equation}
{A_{\mathrm{ani}}^{I}}=\frac{\mu_{0}}{4 \pi} \frac{\gamma_{e} \gamma_{I}}{S} \int \frac{\rho_{s}\left(\mathbf{r}+\mathbf{R}_{I}\right)}{r^{3}} \frac{3 r_{i} r_{j}-\delta_{i j} r^{2}}{r^{2}} d \mathbf{r},
\end{equation}
where $\mu_{0}$ is the permeability of vacuum, $\gamma_{e}$ and $\gamma_{I}$ are the gyromagnetic ratios of electron and nuclei. $\delta_{T}(\mathbf{r})$ is a smeared out $\delta$ function in the relativistic case~\cite{blochl2000}. $\rho_{s}$ is the spin density of spin state $S$ at coordinates $\mathbf{r}$ with respect to the position of the nucleus $\mathbf{R}_{I}$.

The radiative lifetime calculations were performed using an independent-particle approximation approach on the HSE Kohn-Sham states. The transition dipole moment was calculated using the wavefunctions from VASP and the PyVaspwfc code~\cite{Zheng2018}. We use the Wigner-Weisskopf theory of fluorescence to compute the radiative lifetime give by:~\cite{Gali2019,Alkauskas2016,Davidsson2020}

\begin{equation}
\frac{1}{\tau}=\frac{n_{r} (2 \pi)^3 \nu^{3}|\bar{ \boldsymbol{\mu}}|^{2}}{3 \varepsilon_{0} h c^{3}},
\end{equation}
where $\tau$ is the radiative life time, $n_{r}$ is the refractive index of silicon, $\nu$ is the transition frequency, $\bar{\boldsymbol{\mu}}$ is the $k$-point averaged transition dipole moment, $\varepsilon_{0}$ is the vacuum permittivity, $h$ is the Planck constant and $c$ is the speed of light. The transition dipole moment at a single $k$ point is:

\begin{equation}
\boldsymbol{\mu}_{k}=\frac{\mathrm{i} \hbar}{\left(\epsilon_{\mathrm{f}, k}-\epsilon_{\mathrm{i}, k}\right) m}\left\langle\psi_{\mathrm{f}, k}|\mathbf{p}| \psi_{\mathrm{i}, k}\right\rangle,
\end{equation}
where $\epsilon_{\mathrm{i}, k}$ and $\epsilon_{\mathrm{f}, k}$ stands for the eigenvalues of the initial and final states, $m$ is electron mass, $\psi_{\mathrm{i}}$ and $\psi_{\mathrm{i}}$ stands for the initial and final wavefunctions, and $\mathbf{p}$ is the momentum operator.

The chemical potential of hydrogen $\mu_{\rm H}$ can be linked to the synthesis conditions using an ideal gas model.

\begin{equation}
{\mu}_{\rm H}(T,P) = \frac{1}{2}(E_{\rm H_2} - TS_{\rm H_2}^{\rm exp} + RT{\rm ln}(P_{\rm H2})), 
\end{equation}
where $E_{\rm H_2}$ is the energy of the H$_2$ molecule. $S_{\rm H_2}^{\rm exp}$ is the entropy of H$_2$ measured experimentally at standard condition~\cite{Chase1998}, $p_{\rm H2}$ is the partial pressure of H$_2$. We used the H$_2$ molecule at 0K as our reference, zero chemical potential.

\section{Results}

\subsection{Formation energy and charge transition levels}
The T-center is composed of bound carbon atoms substituting on a single silicon site, with a hydrogen atom binding directly to one of the carbon atoms (see Figure 1a). This specific atomic structure of the T-center has been assigned from previous magnetometry, isotope shift and stress studies~\cite{Safonov1996,Safonov1999}. More recent work from Bergeron et al. is consistent with this previously suggested structure~\cite{Bergeron2020}. We used this proposed structure to perform first principles calculations using the hybrid HSE functional. Since defects can exhibit different charged states, we calculated a negatively charged -1 defect and a positively charged +1 defect, in addition to the neutral T-center. Figure 1b shows the defect formation energy versus Fermi level for the T-center in these different charged states. The defect shows a stable neutral state for almost the entire range of Fermi levels within the band gap indicating that doping or gate voltage should not affect the defect charge state. The +1 state is not stable within the band gap and the -1 state is stable only for Fermi levels very close to the conduction band edge. The transition level between the 0 and -1 state is 1.07 eV above the valence band maximum (VBM), 44 meV below the conduction band minimum (CBM).

\begin{figure}[th!]
 	\centering
 	\includegraphics[width=0.45\textwidth]{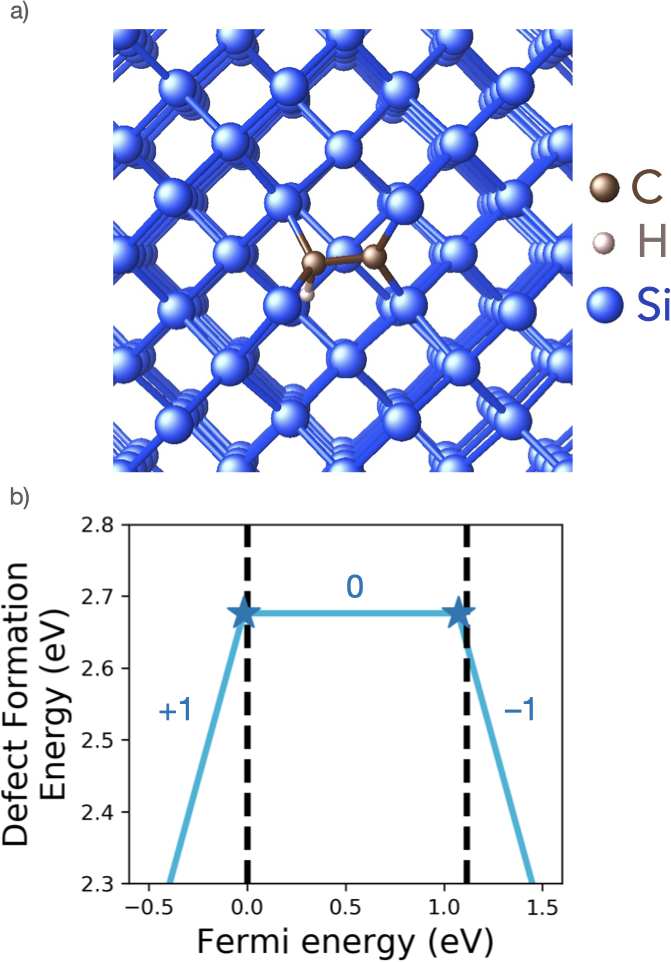}    
 	\caption{(a) Atomic structure of T-center. Two bound carbon atoms replace a single silicon atom. A hydrogen is bound to one of the carbons.  (b) Defect formation energy versus the Fermi energy for T-centers with charges $–$1, 0, and 1. The zero of Fermi energy is referenced to the valence band maximum. We used the chemical potentials of elemental carbon, silicon, and the H$_2$ gas as reference.}
 	\label{Fig.1 T-center defect}
\end{figure}

\subsection{Electronic structure of the ground and excited states}
The ground state, single-electron, Kohn-Sham levels within HSE for the T-center are shown in Figure 2a. The bands from the silicon host are represented by the continuum in blue (valence band) and green (conduction band). The defect levels obtained from our calculations are shown in dark blue for the states due to the silicon host and in red for defect states. Given the point group symmetry of the center, C$_{1h}$~\cite{Safonov1996}, we identify an a$'$ state where both spin up and down levels are degenerate 300 meV below the valence band maximum. A second defect state with symmetry a$''$ is split between spin up and spin down with a large splitting of more than 1 eV. The spin up state is below the valence band maximum while the spin down state is at the CBM (6 meV higher than the CBM). We note that this result is not consistent with earlier \textit{ab initio} calculations where they found the ground state Kohn-Sham levels to have a single spin up and spin down level in the gap with a further defect level in the conduction band~\cite{Leary1998}. Compared to our results, they appear to have systematically higher defect level energies. This is somewhat surprising given that their choice of exchange-correlation functional -- the local density functional -- tends to systematically underestimate fundamental bandgaps and defect level energies\cite{Stampfl_et_al:2000}, which is opposite to the trend we observe from our HSE results. However, several other factors could be causing this discrepancy including their cluster model or their use of a Gaussian basis set (as opposed to our planewave basis set).  Figure 2c shows our calculated isosurface for the square of the wavefunction of the a$''$ state, showing that that it is well localized on the defect. The unpaired electron localizes in an orbital of strong carbon \textit{p} character. The localization of this unpaired electron is reminiscent to the molecular ethyl radical CH$_3$CH$_2\bullet$~\cite{Levine2020,Pacansky1991,Haber2006}. We also computed the hyperfine coupling matrix of the H nuclei with the spin density and found the three eigenvalues for the hyperfine tensor to be 5.39, 4.14 and -2.08 MHz. This is in fair agreement with the experimental data from Bergeron et al. which measured an average hyperfine coupling of 2.6MHz. However, in this work they could only observe the anisotropy of the hyperfine coupling without being able to resolve the full tensor.

\begin{figure}[t]
 	\centering
 	\includegraphics[width=0.45\textwidth]{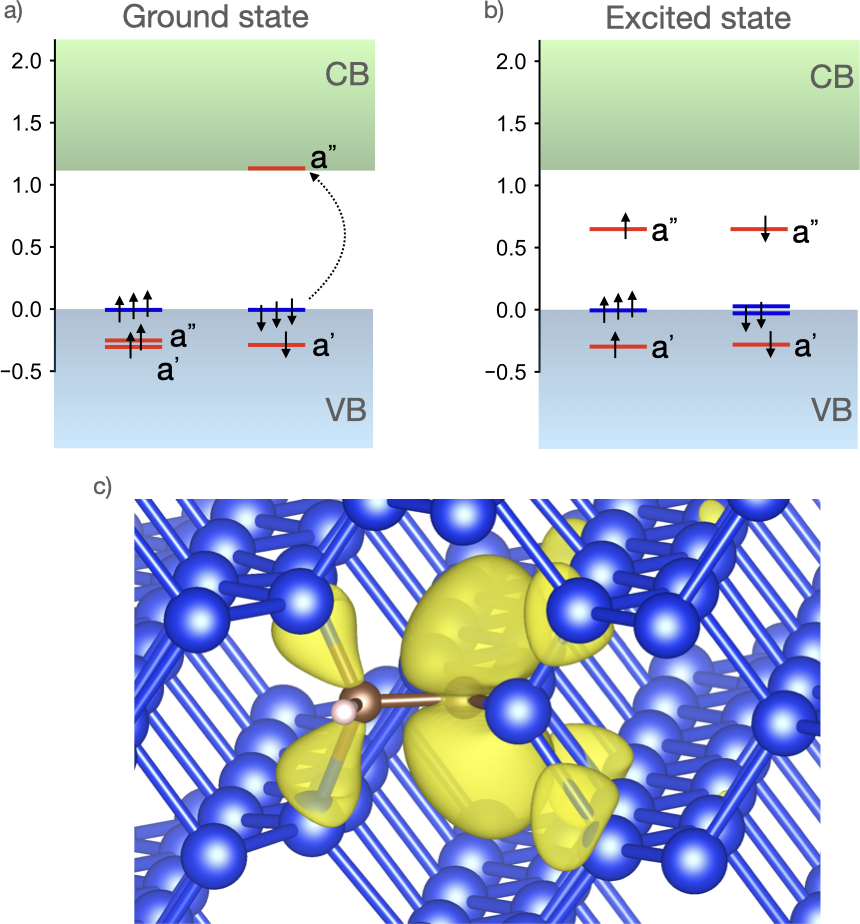}    
 	\caption{Kohn-Sham levels of T-center using HSE in (a) the ground state and (b) the excited state (using constrained-HSE). (c) isosurface for the square of the wavefunction for the a$''$ level.
 	}
 	\label{Fig.2 KS levels}
\end{figure}

We find that the lowest energy excitation is from the valence band edge to the localized a$''$ state. The excited state is made of an electron in a localized state and a hole from a band edge state. Using constrained-HSE, we show in Figure 2b the electronic structure of the defect when the electron is promoted to the localized state leaving a hole in the highest energy band edge state. This hole and electron bind to form a defect-bound exciton. This excited defect can be classified as an isoelectronic acceptor~\cite{Patrick1967}. Our assignment of the electron being in a localized state while the hole is delocalized is in agreement with most previous studies including the recent work from Bergeron et al., but disagrees with an earlier assignment from Irions et al. which assigned the electron to the delocalized state and the hole to the localized one~\cite{Safonov1996,Safonov1999,Irion1985}.

We next calculate the zero-phonon line (ZPL) energy. To calculate the ZPL, we use our constrained-HSE calculations which provide the total energy of the relaxed defect when forcing the occupation of the a$''$ spin down state. The difference between the energy of this excited state and the energy of the ground state is precisely the ZPL energy. While this approach does not fully treat  electron-hole interactions, it does not have the same computational cost as more sophisticated methods (e.g. GW-BSE) and is more amenable to use in a supercell approach. Constrained-HSE has shown good agreement with more advanced theories such as time-dependent density functional theory (TD-DFT) as well as experiments in the case of quantum defects in diamond~\cite{Jin2021}. We calculate a ZPL energy of 985 meV which compares reasonably well with the 935 meV ZPL measured experimentally~\cite{Bergeron2020}.

We now use information from the wavefunction of the ground and excited states to estimate the radiative lifetime. We first calculate the transition dipole in the independent-particle approximation from the HSE electronic structure and find a transition dipole moment of 2.2 D and 0.46 D depending on the use of the excited or ground state electronic structure. The radiative lifetime can then be computed using these transition dipole moments and range from 1.3 to 5 $\mu$s depending on the use of the ground state or excited electronic structure. This is longer than the experimental value of 0.9 $\mu$s reported by Bergeron et al~\cite{Bergeron2020}. Our larger values could indicate a component of non-radiative transition in the experiment.  The long lifetime of the T-center excited state compared to other quantum defects observed in experiments is confirmed by our first-principles calculations. The NV center for instance has a radiative lifetime of around 12 ns both from experiments and calculations with a similar level of theory~\cite{Batalov2008,Siyushev2013}. 

The longer lifetime is inherent to the defect-bound exciton nature of the excited state and to the large difference in nature between the localized and delocalized states constituting this exciton. The transition dipole moment is sensitive to the overlap between the wavefunctions of these two states which are sufficiently different to result in a small overlap. Moreover, since the $sp^3$ hybridization in the bulk now permits the nominally forbidden $p \leftrightarrow p$ transition, this further suggests a small dipole matrix element.  For these reasons, we expect most weakly bound excitons in silicon to show longer radiative lifetime than the NV center and thus lower brightness.
We note that the ground state valence band edge states that are 3-fold degenerate in a perfect bulk silicon supercell are here split by the presence of the defect. This splitting is consistent with the second peak observed experimentally above the ZPL and called TX$_1$ in Ref.~\cite{Bergeron2020}. The splitting between the valence band states in our HSE calculations is 11 meV, which is larger than the experimental value of 1.76 meV. This overestimation likely comes from finite supercell size effects as demonstrated by Zhang et al. on the silicon divacancy in diamond~\cite{Zhang2020}.
Combining the calculated transition levels between the neutral and charge -1 state (0/-1 transition level in Figure 1) of 1.07 eV and the calculated ZPL energy of 985 meV, we can evaluate the energy required to release the weakly bound hole from the defect-bound exciton as 1.07 eV - 0.985 eV=  85 meV. This is larger than the measured activation energy for the thermal photoluminescence (PL) quench (22 meV, 32 meV to 35 meV) ~\cite{Bergeron2020,Safonov1996,Irion1985} but within the error expected from our methodology.

\subsection{Stability versus decomposition and hydrogenation}
In addition to our studies on the stability of the T-center's charge states, we now use first principles computations to assess the stability of the defect against decomposition into simpler defects made from carbon and hydrogen. Table~\ref{tab:table1} shows the reaction energy of decomposition for the T-center (noted as (C-C-H)$\rm_{Si}$) keeping a fixed composition of two carbons and one hydrogen. For this investigation, we used DFT-GGA-PBE rather than the more computationally expensive HSE as the energetics for these processes are already well captured by standard DFT. All decomposition reactions are endothermic indicating that the T-center is thermodynamically stable against decomposition into simpler defects. It has been suggested that the T-center forms through the combination of C substitutional defects and C-H interstitials created by ion implantation~\cite{Safonov1996,Safonov1999}. Our computations agree with this picture and confirm there is a thermodynamic driving force of 1.75 eV for that process. However, other processes could be at play as well as our Table suggests.

\begin{table}[h]
\caption{\label{tab:table1} Reaction energy of decomposing T-center into simpler defects with a fixed compositions of two carbons and one hydrogen}
\begin{ruledtabular}
\begin{tabular}{lc}
\multicolumn{1}{c}{T-center} & T-center\\
\multicolumn{1}{c}{decomposition reaction} & decomposition energy (eV) \\
\hline
(C-C-H)$\rm_{Si}$ $\rightarrow$ 2 C$\rm _{Si}$ + H$_i$ & 2.15 \\
(C-C-H)$\rm_{Si}$ $\rightarrow$ 2 C$_i$ + H$_i$ & 8.41 \\
(C-C-H)$\rm_{Si}$ $\rightarrow$ C$\rm_{Si}$ + C$_i$ + H$_i$ & 5.28 \\
(C-C-H)$\rm_{Si}$ $\rightarrow$ (C-H)$\rm _{Si}$ + C$_i$  & 3.94 \\
(C-C-H)$\rm_{Si}$ $\rightarrow$ C$\rm _{Si}$ + (C-H)$_i$ & 1.75 \\
(C-C-H)$\rm_{Si}$ $\rightarrow$ (C-H)$\rm _{Si}$ + C$\rm _{Si}$ & 0.80 \\
(C-C-H)$\rm_{Si}$ $\rightarrow$ (C-H)$_i$ + C$_i$ & 4.88 \\
(C-C-H)$\rm_{Si}$ $\rightarrow$ (C-C)$\rm_{Si}$ + H$_i$  & 2.81
\end{tabular}
\end{ruledtabular}
\end{table}

The stability of the complex with hydrogenation is also of interest. Given hydrogen's rapid diffusion in solid-state materials, the environment during annealing or operation could lead to hydrogen exchange with the defect. Following the analogy of the T-center with an ethyl radical, it is reasonable to expect that the unpaired electron of the T-center will react spontaneously with hydrogen~\cite{Marshall1999}. In fact, hydrogenation of silicon color centers has been suggested as a potential cause  of PL degradation~\cite{Lightowlers1994}. 
 On the other hand, we also consider a dehydrogenated T-center (C-C)$\rm_{Si}$ which is the ``configuration C" defect previously discussed in the context of the silicon G-center~\cite{Timerkaeva2018}. We use here a thermodynamic model which considers a fixed number of carbon in a silicon matrix (two carbon atoms) and an open system for hydrogen (modeling the environment of the host). The thermodynamic potential is then defined as $E_{\rm defect}-n_{\rm H} \mu_{\rm H}$,  where $E_{\rm defect}$ is the formation energy of the defect obtained from DFT,  $n_{\rm H}$ is the number of hydrogen in the defect, and $\mu_{\rm H}$ the hydrogen chemical potential. Figure 3 shows how the thermodynamic potential of different defects varies with the hydrogen chemical potential. The reference chemical potential ($\mu_{\rm H}$=0 eV) is the H$_2$ molecule at 0K. We include here the T-center with different levels of hydrogenation as well a selection of calculated decomposition products for clarity. All possible decomposition pathways is available in Supplementary Information (see Fig.~S1). When the possibility to exchange hydrogen with the T-center is taken into account, the picture of the T-center stability changes significantly.  The T-center is competing with its hydrogenated ((H-C-C-H)$\rm_{Si}$) and dehydrogenated ((C-C)$\rm_{Si}$) counterparts and will be present only within a narrow chemical potential window (around $\mu_{\rm H}$=-0.92 eV). Within standard conditions (25°C and 1 bar of hydrogen partial pressure, $\mu_{\rm H}$=-0.2 eV), the most stable form of the T-center is the fully hydrogenated one ((H-C-C-H)$\rm_{Si}$). In fact, the ethyl radical molecule also shows a strong tendency to bind with hydrogen under standard conditions~\cite{Marshall1999}. A few typical conditions in terms of temperature and partial pressure of H$_2$ are labeled on Figure 3 (see Methods).

\begin{figure}[th]
 	\centering
 	\includegraphics[width=0.45\textwidth]{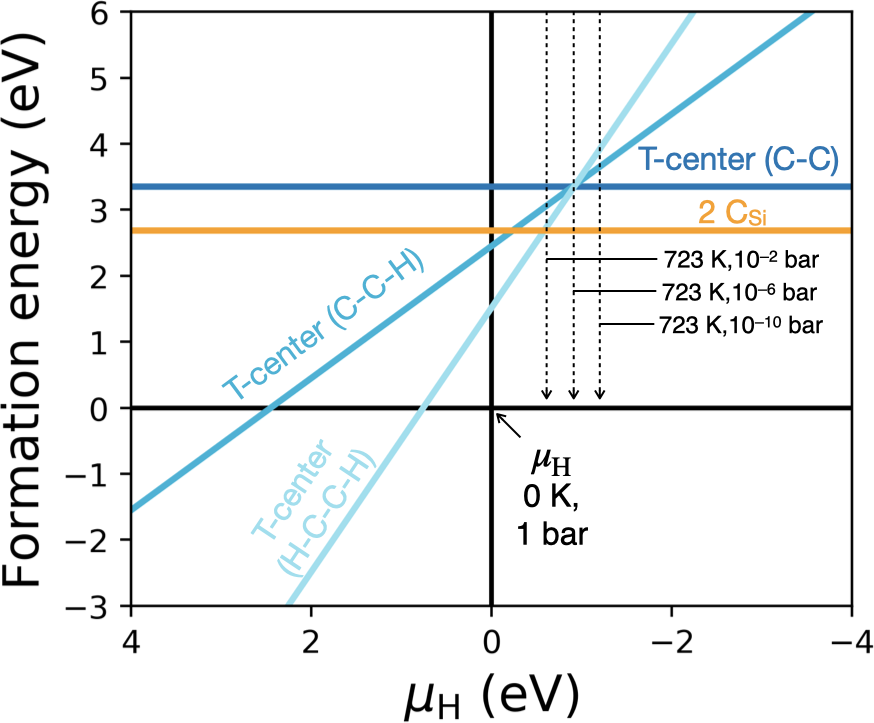}    
 	\caption{Stability of T center in a grand canonical ensemble open to hydrogen}
 	\label{Fig.3 T-center stability}
\end{figure}

Our thermodynamic analysis can be used to rationalize some recent findings in the preparation of the T-center~\cite{MacQuarrie2021}. The preparation steps reported by MacQuarrie et al. starts with carbon implantation and hydrogen implantation. The sample is then annealed under boiling water and finally annealed in nitrogen at higher temperature (around 400°C). Using Figure 3 to guide our understanding, we suggest that the hydrogen implantation and boiling water anneal are there to ensure a high enough hydrogen chemical potential favoring the formation of the hydrogenated form of the T-center (H-C-C-H)$\rm_{Si}$. This agrees with the reported ion implantation procedure of using carbon and hydrogen in a fixed 1:1 ratio~\cite{MacQuarrie2021}. We hypothesize that the high hydrogen chemical potential could be important to favor the formation of the hydrogenated T-center versus the decomposition in 2 substitutional carbon defects (2 C$\rm_{Si}$). The final annealing (at higher temperature and under nitrogen gas) plays a decisive role for successfully forming the T-center. In addition to potentially recovering the ion implanted lattice, our analysis indicates that the annealing is essential to dehydrogenate (H-C-C-H)$\rm_{Si}$ and form the T-center. In view of the narrow range of stability for (C-C-H)$\rm_{Si}$, the annealing conditions (hydrogen partial pressure and temperature) need to be very well controlled. Annealing at too high a hydrogen chemical potential will not bring the driving force necessary for T-center  (C-C-H)$\rm_{Si}$ formation. Annealing at too low a hydrogen chemical potential will favor the fully dehydrogenated T-center (C-C)$\rm_{Si}$. We note that the formation of the 2 carbon substitutional (2C$\rm_{Si}$, orange line in Figure 3) is thermodynamically favored during typical annealing conditions but will be kinetically more difficult than hydrogen exchange and could be kinetically inhibited. Quantitatively, we have found that the optimal annealing conditions for (C-C-H)$\rm_{Si}$ formation lies at $\mu_{\rm H}$=-0.92 eV. This corresponds to a temperature of 450°C and a partial pressure of hydrogen of 10$^{-6}$ bar in good agreement with the experimental observations setting the optimum annealing around 400°C under nitrogen gas. Our analysis also agrees with previous experimental observations indicating that the PL disappears when T-centers are annealed under hydrogen atmosphere as this will lead to the formation of the optically inactive (H-C-C-H)$\rm_{Si}$\cite{Lightowlers1994}. Finally, even when conditions are optimal for the T-center formation, the hydrogenated and dehydrogenated form are very competitive energetically and should be present as well in silicon samples.

\section{Conclusions}

We have reported on a first principles analysis of the T-center in silicon. The defect is stable in the neutral charge state for most of the Fermi levels across the band gap. We compute a ZPL of 985 meV in fair agreement with experiment. The excited state forms a defect-bound exciton by filling a localized state and leaving a bound hole in a valence band state. The localized state shows a strong carbon \textit{p} character reminiscent of the ethyl radical molecule. The bound-defect excitonic nature of the excited state leads to long radiative lifetime on the order of $\mu s$ and low brightness. Finally, we have analyzed the stability of the T-center and rationalize the sequence of implantation such that annealing necessary to its formation. The T-center is stable versus (de)hydrogenation in a very narrow range of hydrogen chemical potential and annealing conditions (atmosphere and temperature) need to be very precisely controlled to lead to a large concentration of T-centers.

\nocite{*}

\section{Acknowledgments}

This work was supported by the U.S. Department of Energy, Office of Science, Basic Energy Sciences in Quantum Information Science under Award Number DE-SC0022289. D.D. acknowledges funding by the Conseil de l’action international (CAI) through a doctorate grant “Coopération au Développement”. This research used resources of the National Energy Research Scientific Computing Center, a DOE Office of Science User Facility supported by the Office of Science of the U.S. Department of Energy under Contract No. DE-AC02-05CH11231 using NERSC award BES-ERCAP0020966. G.H. thanks P. Pochet for useful discussions.

\bibliography{apssamp,alp}

\end{document}


\preprint{APS/123-QED}

\title{Supplemental Information: First principles study of the T-center in Silicon}
\thanks{To whom correspondence should be addressed geoffroy.hautier@dartmouth.edu}%

\author{Diana Dhaliah}
\affiliation{%
 Institute of Condensed Matter and Nanosciences (IMCN), Universit\'{e} Catholique de Louvain, Chemin des \'{E}toiles 8, B-1348 Louvain-la-Neuve, Belgium
}%

\author{Yihuang Xiong}%

\affiliation{%
 Thayer School of Engineering, Dartmouth College, 14 Engineering Dr, Hanover, NH 03755, USA 
}%

\author{Alp Sipahigil}

\affiliation{
 Department of Electrical Engineering and Computer Sciences, University of California,  Berkeley, Berkeley, California 94720, USA
}%
\affiliation{
 Department of Physics, University of California, Berkeley, Berkeley, California 94720, USA
}%
\affiliation{
 Materials Sciences Division, Lawrence Berkeley National Laboratory, Berkeley, California 94720, USA
}%

\author{Sin\'{e}ad M. Griffin}
\affiliation{%
 Materials Sciences Division, Lawrence Berkeley National Laboratory, Berkeley, California 94720, USA
}%
\affiliation{%
 Molecular Foundry, Lawrence Berkeley National Laboratory, Berkeley, California 94720, USA
}%

\author{Geoffroy Hautier}

\affiliation{%
 Thayer School of Engineering, Dartmouth College, 14 Engineering Dr, Hanover, NH 03755, USA
}%
\affiliation{%
 Institute of Condensed Matter and Nanosciences (IMCN), Universit\'{e} Catholique de Louvain, Chemin des \'{E}toiles 8, B-1348 Louvain-la-Neuve, Belgium
}%

\date{\today}

\maketitle

\begin{figure*}[h]
 	\centering
 	\includegraphics[width=0.6\textwidth]{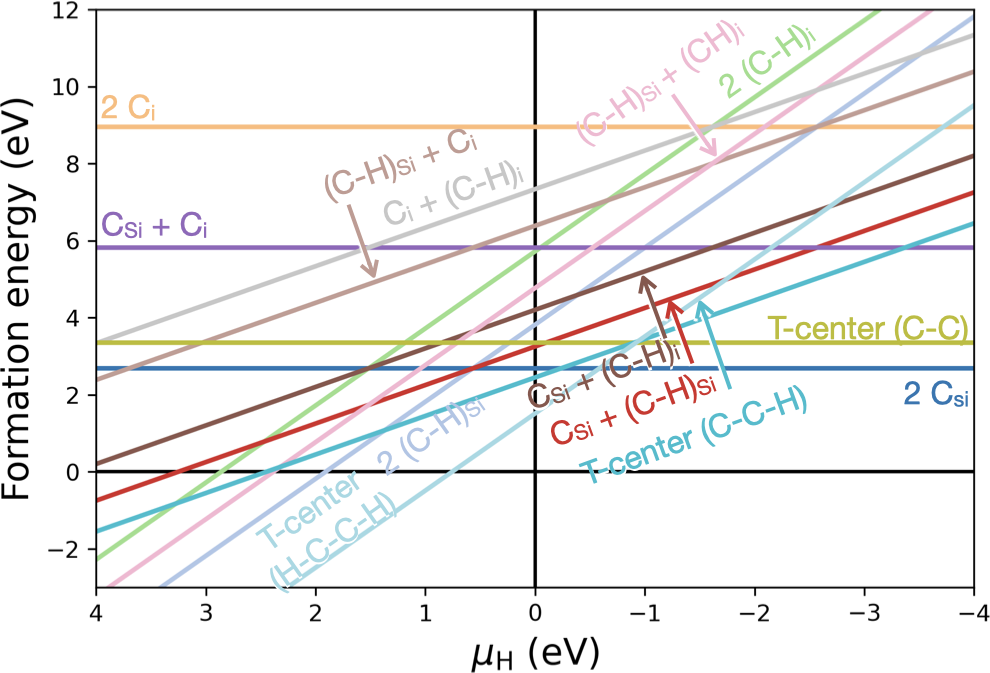}    
 	\caption{The formation energy of all the studied defects with respect to hydrogen chemical potential $\mu_{\rm H}$. Various levels of hydrogenation and dehydrogenation are incoporated to the defect centers, as labeled in the figure.}
 	\label{Fig.3.supp full_figure}
\end{figure*}